\documentclass[11pt,epsfig]{article}
\usepackage{graphicx}

 1
\font\elevenrm=cmr10 scaled\magstep 1

\renewenvironment{thebibliography}[1]
 { \elevenrm
   \begin{list}{\arabic{enumi}.}
    {\usecounter{enumi}     \setlength{\parsep}{0pt}
     \setlength{\itemsep}{3pt} \settowidth{\labelwidth}{#1.}
     \sloppy
    }}{\end{list}}

\begin{document}

\title{{\bf Simple estimates of the masses of pentaquarks with hidden beauty or strangeness}\\}
\author{Vladimir Kopeliovich$^{a,b}$\footnote{{\bf e-mail}: kopelio@inr.ru},
and Irina Potashnikova$^d$\footnote{{\bf e-mail}: irina.potashnikova@usm.cl}
\\
\small{\em a) Institute for Nuclear Research of RAS, Moscow 117312, Russia}
\\
\small{\em b) Moscow Institute of Physics and Technology (MIPT), Dolgoprudny, Moscow district, Russia} 
\\
\small{\em d) Departamento de F\'{\i}sica, Universidad T\'ecnica Federico Santa Mar\'{\i}a;}\\
\small{\it and Centro Cient\'ifico-Tecnol\'ogico de Valpara\'iso,
Avda. Espa\~na, 1680, Valpara\'iso, Chile}
}

\date{}
\maketitle

\begin{abstract}
The masses of cryptoexotic pentaquarks with hidden beauty are estimated phenomenologically using the results 
 by the LHCb collaboration which discovered recently the cryptoexotic pentaquarks with hidden charm.
The expected masses of the  hidden beauty pentaquarks are about $10.8$ GeV and $10.7$ GeV in the limit of some
kind of heavy quark symmetry.                                        
The states with hidden strangeness considered in similar way have masses about $2.37$ Gev and $2.30$ GeV, by several hundreds of MeV higher
than states discussed previously in connection with the relatively light positive strangeness pentaquark $\theta^+$.
Empirical data on spectra of pentaquarks can be used to get information about quarkonia interaction
with nucleons.
The results obtained for the case of heavy flavors are in fair agreement with model of isospin 
(pion) exchange between flavored baryons and anti-flavored vector mesons, proposed by Karliner and Rosner, and in 
qualitative agreement with the bound state version of the chiral soliton model. 
\end{abstract}

\section{Introduction}
 Recent remarkable observation of the pentaquark baryon states with hidden charm and the quark content $c\bar c\,uud$ 
with masses 4450 MeV and 4380 MeV \cite{lhcb1, lhcb2, stone} 
provided new impact for studies  of baryon pentaquark states, which have been almost stopped lately.
The pentaquarks have been observed in \cite{lhcb1} as Breit-Wigner resonances in the system nucleon --- charmonium $J/\psi$. 
This means that there is enough interaction between $J/\psi$-meson and the nucleon, although the branching for the decay 
$P_c \to J/\psi\, p$ was not measured, and other possible decay modes have not been studied. 

Such decay modes are, e.g. $P_c \to \bar D^0 \,\Lambda_c^+$,
or  $P_c \to \bar D^0 \,\Sigma_c^+$,  and  $P_c \to D^- \,\Sigma_c^{++}$, and their branching can be considerably greater than
branching of the decay $P_c \to J/\psi\, p$ which has highly efficient dimuon trigger due to decay $J/\psi \to \mu^+\mu^-$. 
The Breit-Wigner refsonance structure should be seen in these modes, quite similar
to the case of $P_c \to J/\psi\, p$. The LHCb collaboration is going to study these decay modes which contain two charmed 
particles, although they do not expect to see very
many events in these decays with the current data sample \footnote{Private communication by Sheldon Stone, leader of experiments \cite{lhcb1, lhcb2}.}.

The mass of these pentaquark states is mainly due to the masses of heavy charmed quark and antiquark.
A natural question is what could be the masses of analogous cryptoexotic pentaquarks with hidden beauty, which have the
quark content $b\bar b uud$ or $b\bar b udd$. The simplest possible assumption is that such pentaquarks also are similar resonances
in the system $\Upsilon$ - nucleon, and their masses also are mainly due to masses of heavy bottom quark and antiquark.

Here we present a very simple estimate of masses of cryptoexotic pentaquarks with hidden beauty, 
in assumption that the states observed in  \cite{lhcb1} are indeed pentaquarks (not some threshold or another kinematical effect), and that structure
of the hidden beauty pentaquark is similar to the structure of the hidden charm pentaquark \cite{lhcb1}.

Similar estimates can be made also for cryptoexotic pentaquarks with hidden strangeness (in this case the strange quarkonium, or strangeonium, 
is the $\phi (1020)$ meson). The resulting masses are by several hundreds of MeV higher than masses of such states
discussed previously \cite{dpp,wk,aaps,mart}. 
To make these estimates we assume that similaeity of the quarkonia interaction with nucleons takes place. Such similarity of
strangeonium, charmonium and bottomonium has been pointed out in \cite{clm} for the case of meson dynamics.
These estimates, being straightforward and almost trivial, may have nontrivial consequences after observation of the pentaquark states
with hidden flavor, beauty and strangeness. Unique information about quarkonia interactions with nucleons could be extracted from these data.

\section{Masses of pentaquarks from masses of quarkonia}
The simplest estimate of the pentaquarks masses with hidden beauty, i.e. containing 
the $b\bar b$ pair,  is the following (\cite{drama} and comment by the author to this article)
$$M(P_b) = M(P_c) + M(\Upsilon) -  M(J/ \psi), \eqno(1) $$
where
$M(\Upsilon) = 9460$ MeV  is the mass of the lightest bottomonium,
$M(J/\psi) = 3097$ MeV  is the charmonium mass, and we obtain
$$M(P_{b},1) = 10743\, MeV$$
 — the mass of lighter hidden beauty - pentaquark,
$$M(P_{b},2) = 10813 MeV$$ 
— the mass of heavier pentaquark.

This estimate is in the spirit of the heavy quarks symmetry, discussed previously, see e.g. \cite{s-r-r} and references in this paper.
It is supposed here that the role of charmonium and bottomonium in formation of cryptoexotic pentaquarks
is approximately the same, see discussion below.

Similar estimates can be made for the pentaquarks with hidden strangeness.
$$M(P_s) = M(P_c) + M(\phi) -  M(J/\psi),  \eqno(2)$$
where $M(\phi) \simeq 1020$ MeV is the $\phi$-meson mass.  We obtain then 
$$ M(P_s,1) = 2303 \, MeV, \qquad  M(P_s,2) = 2373 \, MeV$$
for the masses of the lower and higher hidden strangeness pentaquarks.

These relations can be written in the form connecting differences of pentaquark masses with differences 
of masses of corresponding quarkonia:
$$ M(P_b) -  M(P_c) = M(\Upsilon) - M(J/\psi); \quad  M(P_c) - M(P_s) = M(J/\psi) - M(\phi). \eqno(3) $$

The contributions of the heavy quark masses obviously satisfy these relations, but do the quark and gluon 
sea contribution satisfy, or not - this is just a question.
At the next step we can include into consideration the difference in the kinetic energies of the motion of 
the pentaquarks constituents - quarkonium and nucleon. We ascribe the difference of pentaquark masses and the quarkonium 
mass plus the nucleon mass to the kinetic energy of the constituents motion.
The reduced mass\footnote{The reduced mass in the system of two particles with masses $m_1,\;m_2$ is 
$m_{12} = m_1m_2/(m_1+m_2)$, the kinetic energy of their relative motion in the center of mass system is
$\vec p ^2/m_1 + \vec p ^2/m_2 =\vec p^2/m_{12}$, i.e. inversely proportional to the reduced mass at fixed
momentum $\vec p$.}
 of the $J/\psi$ meson and nucleon is about 721 MeV, for the $\Upsilon(9460)$ and nucleon it is
854 MeV; for the $\phi$-meson and nucleon it is about 489 MeV.
As a result, the masses of hidden beauty pentaquarks decrease slightly, but masses of hidden  strangeness 
pentaquarks increase by about $\sim 200 \,$MeV, see the table. 

\begin{center}
\begin{tabular}{|l|l|l|l|l|l|l|}
\hline
$ $& $P_c(1) $&$P_c(2)$&$P_b(1)$&$P_b(2)$&$P_s(1)$&$P_s(2)$ \\
\hline
$HQS $         & $4450$(input) & $4380$(input) & $10813$ & $10743$ & $2373$ & $2303$ \\ \hline
$kin.en.corr.$ & $4450$ & $4380$ & $10748$ & $10689$ & $2565$ & $2466$ \\ \hline
\hline
$threshold $ & $4462$ &         & $11139$ & $   $ & $2085 $ & $  $ \\ \hline
\end{tabular}
\end{center}
{\bf Table.} The masses of cryptoexotic states with hidden charm (input, taken from \cite{lhcb1}),
hidden beauty and strangeness. First line of numbers --- the limit of heavy quarks symmetry.
Next line of numbers --- the difference in kinetic energies is taken into account. In the last line the
thresholds are indicated for states consisting of corresponding $\Sigma$-baryon and flavored vector
meson ($\bar D^*(2009)$, $B^*(5325)$, or $\bar K^*(892)$). \\

Note, that masses of cryptoexotic pentaquark states with hidden strangeness obtained in this way are considerably - by several hundreds of MeV -
greater than masses of similar states discussed previously in connection with possible observation of the positive
strangeness pentaquark $\theta^+ (1540)$ \cite{dpp,wk,aaps,mart}\footnote{It should be mentioned here that the name $\theta^+$ has been proposed
by D.I.Diakonov for the low mass positive strangeness pentaquark seen by several collaborations in 2002 - 2004. In calculation \cite{wk}
the mass of $\theta^+(1540)$ has been used as input. The rough estimates of the pentaquark masses, as well as masses of exotic baryon
systems made in \cite{k91} as a function of the "exoticness" number $m$ (number of additiona quark-antiquark pairs) gave the values of the 
$m=1$ pentaquark masses higher by 
few hundreds of MeV - about 1  GeV above the nucleon mass, see Eq. $(6)$, end of section 2 in \cite{k91}.}.
 A review of experimental situation with observation/nonobservation
of this low mass positive strangeness pentaquark is presented e.g. in \cite{hicks},
 where it is called a "mystery". The most recent high statistics experiment 
to search for $\theta^+$ in the reaction $\pi^-p \to K^-X$ \cite{naruki} provided negative result for the mass interval $(1.44 \, - \,1.58)$ GeV. Discussion 
of theoretical status of relatively light pentaquarks within the chiral soliton models can be found in \cite{did, koppent}. 

The difference of the quarkonium mass and twice the mass of lightest meson with corresponding flavor is
$m_\phi - 2m_{K^+} \simeq 32$ MeV for strangeness, $m_{J/\psi} - 2 m_{D^+} \simeq - 642$ MeV for charm, and
$m_\Upsilon - 2 m_{B^+} \simeq - 1098\, $ MeV for beauty. This illustrates the difference of the sea contributions to the
masses of quarkonia for different flavors. We assumed that same differences in the sea of quark and gluons contributions take place
for the masses of pentaquarks with hidden flavor, and this crucial assumption will be checked when the mases of the hidden strangeness
and beauty pentaquarks will be established. 

Further refinements are of interest and possible, in particular, the difference of interactions of different quarkonia with
nucleons may be included into consideration, see e.g. \cite{klz,kaivol,hikt,dos}. This may demand considerable efforts,
because direct measurements of this difference are not possible in view of absence of the quarkonia beams. On the other hand, detection and studies 
of cryptoexotic hidden flavors pentaquarks could provide information on different quarkonia interactions with nucleons, which is difficult
to obtain in other ways.
The comparisons with other approaches, discussed below, are of interest.\\

\section{Comparison with the molecular type models} 
A natural way to estimate the masses of pentaquarks is to consider them as dynamically generated in meson-baryon interactions \cite{wm}
or as a molecular-type bound states of
baryons and mesons, as it was made in \cite{ys,xo,kr-1,hdpw}. Several states with hidden charm and masses above $4\,GeV$ have been predicted in
\cite{wm} where the decay channel $P_c \to J/\psi\,p$ also has been pointed out as convenient for detection of this particle. 
As in our case, for heavy flavors the relative accuracy of such estimates is
better than for lighter flavors. The hidden charm states have been obtained as $\Lambda_c$ (or $\Sigma_c$) --- $\bar D\, (\bar D^*)$ molecules, 
and similar for hidden bottom states.
 This approach allowed to make more detailed predictions of binding energies, and it is not in contradiction with
our simple estimates. Several states with hidden beauty and masses around $11\, GeV$ have been obtained in \cite{xo} as result of meson $(B,B^*)$
baryon $(\Lambda_b,\, \Sigma_b,\,\Sigma^*_b)$ interactions due to exchange of vector mesons.

  
The isovector meson (pion) exchange binding mechanism has been proposed in \cite{kr-1}, which could provide binding of the $\Sigma_c \, \bar D^*$ state
(threshold energy $4462.4 $ MeV) in fair agreement with observation of \cite{lhcb1}. There are no bound states containing $\Lambda_c$ 
or $\Lambda_b$ in this approach because of isospin properties of the pion exchange.
Similar estimates for hidden strangeness pentaquarks would be of interest.
The molecular-type hidden strangeness pentaquarks consisting of $\Sigma$-hyperon and $\bar K^*$ meson should have the mass not greater
than corresponding threshold, about $2085$ MeV, also greater than results obtained previously within the chiral soliton
approach \cite{dpp, wk}, but smaller than estimates made above. Apparently, this can be connected with the fact that
the mass of the $\phi$-meson is greater than the double kaon mass, whereas the mass of the $J/\psi$-meson is by more than six hundreds of MeV
smaller than double $D$-meson mass, and the mass of the $\Upsilon$-meson is smaller than double $B-$meson mass by more than $1\;GeV$.

To make the whole picture of pentaquarks more complete, it is important to find the states with exotic quantum numbers --- positive strangeness
(discussed in \cite{dpp,wk,hicks}), negative charm, positive beauty. Such states could belong to some $SU(3)$ multiplets together with the
states with hidden flavor. If we are building the pentaquark states from the octet of baryons and the nonet of vector mesons, we shall obtain
antidecuplet and $\{27\}$-plet of pentaquarks. The isospin zero $\theta^+ \in \overline{10} = (K^{*+} n -K^{*0}p)/\sqrt 2$,
$\theta^{*++}\in \{27\} = K^{*+}\,p$, etc. The masses of these states are expected to be smaller than threshold energy $\sim 1830$ MeV.

\section{Remarks on the bound state chiral soliton model} 
As we mentioned already, dramatic events took place around observation/nonobservation of the relatively light positive strangeness
pentaquark predicted just within a variant of the chiral soliton model in \cite{dpp}. Several years yearlier the crude estimate
of the pentaquark masses provided the values about $1$ GeV above the nucleon mass \cite{k91}, even without the flavor symmetry
breaking mass terma contributions.
Here we show that phenomenological estimates we made in section 2 are in qualitative agreement with the bound state version of the chiral soliton model,
first proposed for description of exotic heavy flavored states in \cite{rs93} and recently discussed in \cite{s-r-r}.

The chiral soliton approach is conceptually different from approaches discussed in previous section, because each 
state is characterized by external quantum numbers, but the constituents (building blocks) of the state do not come into play at all.
In such models the flavor/antiflavor excitation energies are equal \cite{klw1,klw2,kopshun}
$$ \omega_F = {N_c\over 8\Theta_F} (\mu_F - 1) ; \quad \bar \omega_{\bar F} = {N_c\over 8\Theta_F} (\mu_F + 1) \eqno(4) $$
where $N_c$ is the number of colors of the underlying QCD, $\Theta_F$ is the so called flavored moment of inertia of skyrmion, 
$\Gamma $ is proportional to the $\Sigma -$ term of the
nucleon (it is the "rigid oscillator" version of the model proposed by Klebanov and Westerberg \cite{klw1,klw2}; definitions and slightly
modified complete formulas can be found in \cite{kopshun} and references in this paper).
$$\mu_F = {1\over N_c} \left[ N_c^2 + 16 \Theta_F\left(\bar m_D^2 \Gamma +(F_D^2 - F_\pi^2) \tilde \Gamma\right) \right]^{1/2} \eqno(5) $$
is dimensionless quantity with
$$\bar m_D^2 = {F_D^2\over F_\pi^2} m_D^2 - m_\pi^2. \eqno(6) $$
$m_D$ is the $D-$ meson (or $B-$ meson, or $K$-meson) mass, $F_D$ is the corresponding meson decay constant, e.g. $F_K/F_\pi \simeq 1.22$,
$F_D/F_\pi \simeq 1.58$ \cite{rs}.

At large eenough mass $m_D$  we have approximately
$$ \mu_F \simeq 4 \bar m_D {\sqrt{\Gamma\Theta_F}\over N_c} \eqno(7) $$
and
$$\omega_F \simeq {F_D\over F_\pi} m_D \sqrt{{\Gamma\over 4\Theta_F}} \eqno(8)$$
Since
we have always $\Gamma/4\Theta_F <1$  \cite{kopshun}, the quantity
$$ \epsilon_F= m_D\left[1 - {F_D\over F_\pi}\sqrt{{\Gamma\over 4\Theta_F}}\right] \eqno(9)$$
can be interpreted as a binding energy of the heavy meson by skyrmion.

For cryptoexotic states the sum enters
$$\omega_F + \omega_{\bar F} = {N_c\over 4\Theta_F} \mu_F \simeq {F_D\over F_\pi} m_D \sqrt{{\Gamma\over \Theta_F}} \eqno(10) $$
which defines the main contribution to the difference of masses of the pentaquark and nucleon. 
Relations $(7)-(10)$ work better for heavy flavors, charm or beauty, for the case of charm  $(10)$ gives the value about $3.31$ GeV,
for reasonable choice of the model parameters, see \cite{kopshun},
in  agreement with data \cite{lhcb1}, and somewhat greater than the mass of the $J/\psi$-quarkonium. 
For beauty the knowledge of the
ratio $F_B/F_\pi$ is lacking still. If we take this ratio equal to that for charm, we obtain from $(10)$ the value $9.35$ GeV
\footnote{For these estimates we used the values $\Gamma=4.83\,GeV^{-1}$, $\tilde \Gamma =15.6\, GeV^-1$, $\Theta_F^0 = 2.04\,GeV^{-1}$,
$\Theta_c =\Theta_F^0+(F_D^2/F_\pi^2 - 1)\Gamma/4 = 3.85\,GeV^{-1}$,  see\cite{kopshun} and references in this paper.},
which gives the mass of the $P_b$ pentaquark abour $10.29$ GeV (lower boundary), in fair agreement with estimates of section 2.

For strangeness the result
from Eq. $(5)$ for the sum of energies $(10)$ is about $0.84$ GeV, by $0.18$ GeV lower than the mass of the $\phi$-meson. 
It is lower than estimate made above in section 2, and close to previous estimate \cite{k91}, but spin and isospin dependent
corrections should be included as well.
Relations $(9),\,(10)$ are of interest because they connect quantities of differen nature: flavor decay constants $F_D,\,F_\pi$,
skyrmions characteristics $\Gamma,\;\Theta_F$ 
and the masses of hadrons. Spin and isospin dependent hyperfine splitting correction to the energy of the state should be included
for more detailed comparison with data \cite{klw1,kopshun}, this will be done elsewhere.

Within the chiral soliton approach the exotic (also nonexotic) states naturally belong to definite $SU(3)$ multiplets of baryons ---
antidecuplet \cite{dpp}, $\{27\}-$plet or $\{35\}-$plet \cite{wk}.  The states we discuss here most probably are the partners of the lowest
states which belong to definite $SU(3)$ multiplets, discussed in section 7 of \cite{koppent}b. They can be some mixtures of the components
of different $SU(3)$ multiplets, and this point needs further clarification.
\\

\section{Conclusions and prospects} 
We have estimated the masses of pentaquarks with hidden flavor, beauty and strangeness,
using a simplified phenomenological model of the bound state of quarkonium and nucleon.
These estimates may be useful for planning future experiments and as starting point of
more refined study. The spectra of such states will be a source of information about quarkonia --- nucleons interaction
and about flavor dependence of the quarks and gluon sea contribution to the pentaquark masses. 
In view of uncertainties intrinsic to the model and the way of calculation
we pretend on the qualitative agreement with data, only. 

The results obtained within the bound state version of the chiral soliton model (section 4) are in reasonable agreement 
with data for the hidden charm pentaquark \cite{lhcb1} and support the phenomenological estimate of section 2 for the 
mass of hidden beauty pentaquark. It would be important and very interesting to find and study manifestly exotic baryon
states, i.e. with negative charm, or positive strangeness or beauty, to complete the picture of baryons exotics.

The pentaquark states with hidden flavors appear naturally in the molecular-type models where the pentaquark consists of
flavored baryon and anti-flavored meson, vector \cite{xo} or pseudoscalar \cite{kr-1,hdpw}. The point is that in 
isospin exchange models \cite{kr-1}, and in our estimates the masses of hidden 
strangeness states  turn out to be higher than masses of states discussed previously in connction with 
supposed existence of the relatively light positive strangeness pentaquark $\theta^+$ \cite{dpp,wk,aaps,mart}.

Few days after this paper has been submitted to HEP database (arXiv:1510.05958 [hep-ph]), the paper \cite{lebed} 
appeared where the possibility to detect the 
pentaquarks with hidden strangeness in decays of $\Lambda_c(2286)\to P_s^+\pi^0 \to \phi p\pi^0 $, which is analogous to the
discovery channel of $P_c^+$ \cite{lhcb1}, $\Lambda_b \to P_c^+ K^- \to J/\psi\, p\,K^-$, was pointed out. This possibility could be realized
only if the mass of $P_s^+$ is low enough, smaller than $\sim 2151\, $ MeV. The possibilities to detect the states with masses $2372$ MeV 
and $2303$ MeV, indicated here after Eq. (2) and in the table above, have been discussed in version 3 of paper \cite{lebed}. 

We are indebted to Boris Kopeliovich for his comments on the quarkonia interactions with nucleons, and useful remarks. VK is grateful to Prof. 
Sheldon Stone for detailed information
about plans of LHCb to study decay channels of pentaquarks $P_c$.
This work is supported by Fondecyt (Chile), grant number 1130549.
\\


{\bf References}

\baselineskip=11pt
\begin{thebibliography}{99}


\bibitem{lhcb1} LHCb Collaboration (Roel Aaij et al.) Observation of $J/\psi\,p$ resonances consistent with pentaquark states in 
$\Lambda^0_b\to  J/\psi K^- p$ decays.  Phys.Rev.Lett. 115 (2015) 072001; e-Print: arXiv:1507.03414 [hep-ex].

\bibitem{lhcb2} LHCb Collaboration (Roel Aaij et al.). Study of the production of $\Lambda^0_b$ and $B^0$ hadrons 
in $pp$ collisions and first measurement of the $\Lambda^0_b \to J/\psi p\, K^-$ branching fraction.\\ 
CERN-PH-EP-2015-223, LHCB-PAPER-2015-032 
e-Print: arXiv:1509.00292 [hep-ex]

\bibitem{stone} Sheldon Stone.	Pentaquarks and Tetraquarks at LHCb 
Conference: C15-07-22, Conference: C15-08-04 Proceedings.
arXiv:1509.04051 [hep-ex] 

\bibitem{dpp} Dmitri Diakonov, Victor Petrov, Maxim Polyakov. 
Exotic anti-decuplet of baryons: Prediction from chiral solitons 
Z.Phys. A359 (1997) 305;  arxiv: hep-ph/9703373

\bibitem{wk} H.~Walliser, V.B.~Kopeliovich. Exotic baryon states in topological soliton models.
J.Exp.Theor.Phys. 97 (2003) 433 [Zh.Eksp.Teor.Fiz. 124 (2003) 483];
arxiv: hep-ph/0304058 
	
\bibitem{aaps} 
R.A.~Arndt, Ya.I.~Azimov, M.V.~Polyakov, I.I.~Strakovsky, R.L.~Workman. 
Nonstrange and other unitarity partners of the exotic Theta$^+$ baryon.  Phys.Rev. C69 (2004) 035208;
arxiv: nucl-th/0312126

\bibitem{mart} 
T.~Mart. Evidence for the $J^P=1/2^+$ narrow state at 1650 MeV in the photoproduction of $K\,Lambda$.  
Phys.Rev. D83 (2011) 094015; 
arXiv:1104.2389 [hep-ph]


\bibitem{clm} Dian-Yong Chen, Xiang Liu, Takayuki Matsuki.
Two Charged Strangeonium-Like Structures Observable in the $Y(2175)\to \phi(1020)\pi^+\pi^-$ Process.
Eur.Phys.J. C72 (2012) 2008; 
arXiv:1112.3773 [hep-ph

\bibitem{drama} V.~Kopeliovich. Continuation of the pentaquarks drama.
http://trv-science.ru/2015/07/28/prodolzhenie-pentaquarkovoj-dramy/

\bibitem{s-r-r} N.N.~Scoccola, D.~Riska and M.~Rho.	
Pentaquark candidates $P^+_c(4380)$ and $P^+_c(4450)$ within the soliton picture of baryons 
Phys.Rev. D92 (2015) 5, 051501;
arXiv:1508.01172 [hep-ph] 

\bibitem{k91} V.~Kopeliovich. 
On exotic systems of baryons in chiral soliton models.
Phys.Lett. B259 (1991) 234; electronic version is available as arXiv:1601.06493 [nucl-th] 

\bibitem{hicks} K.H.~Hicks. On the conundrum of the pentaquark. Eur. Phys. J. H37 (2012) 1.
	
\bibitem{naruki} M.~Naruki for the J-PARC E19 Collaboration. 
Search for Pentaquark Theta+ in Hadronic Reaction at J-PARC.
e-Print: arXiv:1602.03951 [nucl-ex]

\bibitem{did} Dmitri Diakonov. From pions to pentaquarks.  
arXiv: hep-ph/0406043

\bibitem{koppent} V.B.~Kopeliovich.
Exotic baryon resonances and the model of chiral solitons  
Phys.Usp. 47 (2004) 309 [Usp.Fiz.Nauk 174 (2004) 323];
Pentaquarks in chiral soliton models: notes and discussion. Phys.Part.Nucl. 37 (2006) 623 [Fiz.Elem.Chast.Atom.Yadra 37 (2006) 1184];
arXiv: hep-ph/0507028

\bibitem{klz} B.Z.~Kopeliovich, L.I.~Lapidus, A.B.~Zamolodchikov. Dynamics of Color in Hadron Diffraction on Nuclei.   
JETP Lett. 33 (1981) 595 [Pisma Zh.Eksp.Teor.Fiz. 33 (1981) 612]

\bibitem{kaivol} Heavy quarkonia interactions with nucleons and nuclei 
A.B.~Kaidalov, P.E.~Volkovitsky.  Phys.Rev.Lett. 69 (1992) 3155


\bibitem{hikt} J.~Hufner, Yu.P.~Ivanov, B.Z.~Kopeliovich and A.V.~Tarasov. Photoproduction of charmonia and total charmonium proton cross-sections. 
Phys.Rev. D62 (2000) 094022 
	


\bibitem{dos}  
I.V.~Danilkin, V.D.~Orlovsky, Yu.A.~Simonov. Hadron interaction with heavy quarkonia 
Phys.Rev. D85 (2012) 034012; e-Print: arXiv:1106.1552 [hep-ph] 





\bibitem{wm} Jia-Jun Wu, R. Molina, E. Oset, B.S. Zou. 
Prediction of narrow $N^*$ and $\Lambda^*$ resonances with hidden charm above 4 GeV 
Phys.Rev.Lett. 105 (2010) 232001;
 arXiv:1007.0573 [nucl-th] 

\bibitem{ys}  Z.C.~Yang, Z.F.~Sun, J.~He, X.~Liu and S.L.~Zhu, 
The possible hidden-charm molecular baryons composed of anti-charmed meson and charmed baryon.
Chin. Phys. C 36, 6 (2012)
arXiv:1105.2901 [hep-ph] 

	
\bibitem{xo} C.W.~Xiao, E.~Oset. 
Hidden beauty baryon states in the local hidden gauge approach with heavy quark spin symmetry 
 Eur.Phys.J. A49 (2013) 139;
arXiv:1305.0786

\bibitem{kr-1} M.~Karliner, J.L.~Rosner 
New Exotic Meson and Baryon Resonances from Doubly-Heavy Hadronic Molecules.
Phys.Rev.Lett. 115 (2015) 12, 122001;
arXiv:1506.06386	

\bibitem{hdpw} Hongxia Huang, Chengrong Deng, Jialun Ping, Fan Wang. 
Possible pentaquarks with heavy quarks. 
arXiv:1510.04648 [hep-ph]

\bibitem{rs93} D.O.~Riska and N.N.~Scoccola. Anti-charm and anti-bottom hyperons. 
Phys.Lett. B299 (1993) 338


\bibitem{klw1} K.M.~Westerberg and I.R.~Klebanov. 
On hyperfine splittings of strange baryons in the Skyrme model.
Phys.Rev. D50 (1994) 5834; 
arxiv: hep-ph/9406383

\bibitem{klw2} I.R.~Klebanov and K.M.~Westerberg. 
A simple description of strange dibaryons in the Skyrme model. 
Phys.Rev. D53 (1996) 2804; 
arxiv: hep-ph/9508279


\bibitem{rs} J.L.~Rosner and S.~Stone.
Leptonic Decays of Charged Pseudoscalar Mesons.
arXiv:1002.1655 [hep-ex] 

\bibitem{kopshun} V.B.~Kopeliovich and A.M.~Shunderuk. 	
Flavored exotic multibaryons and hypernuclei in topological soliton models.  
J.Exp.Theor.Phys. 100 (2005) 929 [Zh.Eksp.Teor.Fiz. 127 (2005) 1055];
arxiv:  nucl-th/0409010


\bibitem{lebed} Richard F.~Lebed. Do the $P^+_c$ Pentaquarks Have Strange Siblings? 
 Phys.Rev. D92 (2015) 11, 114030;  arXiv:1510.06648 [hep-ph]

\end{thebibliography}
\end{document}